\documentclass[pdflatex,sn-mathphys-num]{sn-jnl}% Math and Physical Sciences Numbered Reference Style
\usepackage{graphicx}%
\usepackage{multirow}%
\usepackage{amsmath,amssymb,amsfonts}%
\usepackage{amsthm}%
\usepackage{mathrsfs}%
\usepackage[title]{appendix}%
\usepackage{xcolor}%
\usepackage{textcomp}%
\usepackage{manyfoot}%
\usepackage{booktabs}%
\usepackage{lineno}
\usepackage{algorithm}%
\usepackage{algorithmicx}%
\usepackage{algpseudocode}%
\usepackage{listings}%
\theoremstyle{thmstyleone}%
\theoremstyle{thmstyletwo}%

\theoremstyle{thmstylethree}%

\begin{document}

\title[ANADEF: A Nested-Permutation Alarm for Dual-Parameter Earthquake Forecasting]{ANADEF: A Nested-Permutation Alarm for Dual-Parameter Earthquake Forecasting}


\author*[1,2]{\fnm{Hamzeh} \sur{Mohammadigheymasi}}
\author[3]{\fnm{Muhammed Hossein} \sur{Mousavi}}

\author[1]{\fnm{Nuno} \sur{Pombo}}\email{hamzeh@ubi.pt}

\affil[1]{\orgdiv{Department of Computer Sciences}, \orgname{University of Beira Interior}, \orgaddress{\city{Covilh\~a}, \country{Portugal}}}

\affil[2]{\orgdiv{Atmosphere and Ocean Research Institute (AORI)}, \orgname{the University of Tokyo}, \orgaddress{\city{Tokyo}, \country{Japan}}}

\affil[3]{\orgdiv{Department of Physics Education}, \orgname{Farhangian University}, \orgaddress{\postcode{P.O. Box 14665-889}, \city{Tehran}, \country{Iran}}}

%\linenumbers
\abstract{Spatially resolved stress proxies and rate-based seismicity models are increasingly combined for regional earthquake forecasting, yet formally establishing that such predictors carry non-redundant information, rather than merely improving in-sample fit, remains a largely unaddressed statistical problem. Here we present the Nested-Permutation Alarm for Dual-Parameter Earthquake Forecasting (ANADEF) pipeline, an alarm-based spatial forecasting framework that integrates a stress-sensitive Gutenberg--Richter $b$-value field, estimated via penalized two-dimensional B-spline inversion, with a stationary background seismicity rate ($\mu$) derived from space--time Epidemic-Type Aftershock Sequence (ETAS) stochastic declustering. The framework is evaluated on the Zagros Fold--Thrust Belt using a homogenized, 18-year regional catalog ($n = 40{,}731$ events, $M_{\mathrm{N}} \geq 1.5$, 2006--2024), under a two-stage protocol with strictly non-overlapping training (2006--2014) and target (2015--2024) windows. For $M_w \geq 5.0$ ($N = 55$), the integrated model achieves a retrospective Area Skill Score of $S = 0.69$ (95\% CI: 0.64--0.73), reducing the alarmed area fraction from $\tau \approx 0.38$ under the $\mu$-only benchmark to $\tau \approx 0.28$ while retaining a hit rate of $\nu = 92.7\%$; because the operating thresholds ($q_b^{*} = 0.58$, $q_\mu^{*} = 0.36$; $b_{\mathrm{opt}} = 0.76$, $\mu_{\mathrm{opt}} = 0.038\ \mathrm{events\ deg^{-2}\,yr^{-1}}$) are calibrated on the same evaluation catalog, these figures represent retrospective, in-sample skill rather than independent out-of-sample performance. To test whether the $b$-value field contributes information beyond $\mu$ alone, we develop a nested permutation procedure that re-optimizes decision thresholds within each null realization, thereby explicitly absorbing the optimization bias inherent to threshold calibration. Evidence for incremental $b$-value information is sensitive to the choice of null model: a cell-wise randomization null yields nominal significance ($p = 0.012$), whereas a more conservative spatial-structure-preserving null yields weaker, non-significant evidence ($p = 0.057$), indicating that the incremental contribution of stress-based information over the historical rate benchmark, while suggestive, is not unambiguously established at this sample size. The calibrated decision boundaries were subsequently frozen and applied without modification to updated 2015--2024 predictor fields to generate an unvalidated, forward-looking spatial alarm susceptibility template for 2025--2029. Taken together, these results establish a reproducible, statistically transparent methodology for testing, rather than assuming, the complementarity of stress-sensitive and rate-based seismicity predictors, and provide a candidate operational template for regional hazard characterization pending independent prospective validation.}




\maketitle

\section{Introduction}\label{sec1}

Forecasting the location of future earthquakes remains one of the central open problems in seismology. The irregular recurrence of large-magnitude events, combined with observational catalogs that are short relative to tectonic loading timescales, limits the extent to which predictive models of seismic activity can be independently validated \cite{bakun1985parkfield,schorlemmer2005variations}. A range of approaches has been proposed to characterize spatial seismic hazard, including frameworks based on seismic gaps, stress transfer, pattern recognition, and statistical analysis of earthquake catalogs \cite{dodge1996detailed,king1994static,brodsky2006long,deng1996triggering,gomberg1996stress,pollitz1997kobe}. Among these, Operational Earthquake Forecasting (OEF) provides time-dependent probability estimates by combining statistical point-process models with continuously updated catalogs, most commonly through the Epidemic-Type Aftershock Sequence (ETAS) and Short-Term Earthquake Probability (STEP) models \cite{darzi2022calibration,ebrahimian2022improvements}. Because each earthquake elevates the short-term probability of subsequent events, separating transient, triggered seismicity (foreshocks and aftershocks) from the underlying stationary background process is a prerequisite for any analysis of long-term regional seismicity. Traditionally, this background process has been treated as a spatially uniform, time-invariant Poisson rate; however, catalog-based studies show that background seismicity is itself spatially heterogeneous, motivating declustering methods that recover a spatially resolved, stationary background-rate field rather than a single scalar rate \cite{li2019space,zaliapin2013earthquake}. The present study makes use of this stationary output of the ETAS decomposition, the background rate field $\mu(x,y)$, as a static spatial predictor; it does not implement, and should not be read as, a time-dependent OEF system.

A second, physically distinct spatial predictor is available through the frequency--magnitude distribution of the Gutenberg--Richter relation, parameterized by the $b$-value. Laboratory experiments, numerical simulations, and field observations indicate that low $b$-values correlate with elevated differential shear stress on a fault segment and often precede large ruptures, making spatial $b$-value mapping a standard proxy for identifying locally stressed, hazardous structures \cite{nanjo2012decade,shi2018decrease,scholz2015stress,beall2022linking}. The background rate $\mu$, by contrast, is grounded in rate-and-state friction theory, where elevated background rates reflect sustained tectonic stressing integrated over the catalog history rather than the instantaneous stress state of a specific fault patch \cite{dieterich1994constitutive,zhuang2005variable,ogata2022prediction}. Although both quantities are frequently described in the literature as stress-related proxies, they are sensitive to physically distinct processes operating at different spatial and temporal scales: $b(x,y)$ is a local, near-instantaneous indicator of differential stress on individual fault segments, whereas $\mu(x,y)$ integrates the cumulative, long-term seismic response of a structure to tectonic loading. This distinction implies that the two fields are not a priori redundant, but their apparent complementarity has rarely been tested formally rather than assumed - a gap this study is designed to address directly.

The Zagros Fold--Thrust Belt is one of the most seismically active intracontinental collision zones globally, with seismicity partitioned between the sedimentary cover and the crystalline basement \cite{talebian2004aftershock,nissen2011geology,dolatabadi2021combined,mousavi2026cross,tavakolizadeh2025fqsha}. Despite extensive seismotectonic characterization of the region, the joint spatial structure of $b$-value and background-rate fields has not been examined there. A comparable dual-parameter approach was recently applied to earthquake forecasting in Yunnan Province, China \cite{zhang2025earthquake}, demonstrating the feasibility of combining these predictors operationally; the present study extends that framework to the Zagros and introduces three methodological refinements aimed at strengthening the statistical defensibility of the resulting forecasts: (i) spatial $b(x,y)$ is estimated via penalized two-dimensional B-spline inversion (HIST-PPM) rather than gridded maximum-likelihood estimation, yielding a continuously varying, roughness-regularized field with an associated posterior covariance rather than a piecewise-constant grid estimate; (ii) model comparison is embedded within an explicit four-level null hierarchy (Levels 0--3, Section~\ref{sec2}) that isolates the incremental information contributed by $b$-value stress physics specifically over and above the historical background-rate benchmark, rather than over an uninformative uniform baseline alone; and (iii) a two-stage protocol with strictly frozen, non-overlapping calibration and prospective windows is used together with a cosine-latitude area correction, so that reported alarm fractions are not distorted by the region's latitudinal extent. Of these, (ii) constitutes the primary methodological contribution, as it is what permits the incremental value of $b$-value information to be tested rather than assumed; (i) and (iii) are supporting refinements that improve, respectively, the statistical quality of the $b$-value field and the metric used to evaluate it.

The central research question addressed in this study is therefore narrowly defined: \textit{does the spatially resolved $b$-value field provide statistically significant predictive information beyond that already contained in the background seismicity rate $\mu(x,y)$, once the systematic optimization bias introduced by empirical threshold calibration has been explicitly accounted for?} This question is formalized as a null hypothesis, that the spatial pattern of $b(x,y)$ carries no true association with future large-earthquake locations beyond what $\mu(x,y)$ already predicts, which is tested using a nested permutation procedure (Section~\ref{sec4}) rather than inferred from point-estimate skill scores alone.

To address this question, we develop the Nested-Permutation Alarm for Dual-Parameter Earthquake Forecasting (ANADEF) pipeline, which combines the penalized B-spline $b$-value field with the ETAS-derived background rate $\mu(x,y)$ within a Molchan error-diagram framework, and apply it to a high-density, 18-year earthquake catalog from the Zagros Fold--Thrust Belt ($N=40{,}731$ events, $M_{\mathrm N}\geq1.5$, 2006--2024). Predictor fields are estimated from a training window (2006--2014) and evaluated against an independent target catalog (2015--2024) under a strict two-stage protocol; because the operating thresholds reported in Section~\ref{sec4} are themselves calibrated by maximizing skill on this same 2015--2024 target catalog, the resulting performance metrics are retrospective, in-sample calibrated skill estimates rather than independent out-of-sample forecasts, a distinction we maintain throughout the presentation of results. The calibrated thresholds are subsequently frozen without further tuning and applied to updated 2015--2024 predictor fields to generate an unvalidated, forward-looking spatial alarm susceptibility template for 2025--2029, whose predictive value can only be assessed against future observations. In this sense, the contribution of the present study is twofold: a statistically explicit test of whether stress-sensitive and rate-based seismicity predictors carry non-redundant spatial information in a tectonically complex collision zone, and a reproducible computational pipeline through which that test, and the resulting prospective alarm template, can be extended to other regions.

\section{Methodology}\label{sec2}

\subsection{Spatial \texorpdfstring{$b$}{b}-Value Estimation}\label{sec2.1}
The frequency--magnitude distribution of earthquakes is conventionally described by the Gutenberg--Richter (G--R) relation \cite{gutenberg1944frequency}:
\begin{equation}
\log_{10} N = a - bM \label{eq1}
\end{equation}
where $N$ denotes the cumulative number of earthquakes with magnitude equal to or greater than $M$, $a$ quantifies the overall level of seismic productivity within the study region, and $b$ (the $b$-value) characterizes the relative proportion of large to small events. Under this relation, the continuous probability density function (PDF) of earthquake magnitudes above a completeness threshold $M_c$ is defined by:
\begin{equation}
f(M) = \frac{N(M)}{\int_{M_c}^{\infty} N(M)\,dM} = \beta\,e^{-\beta(M - M_c)}, \quad M \ge M_c \label{eq2}
\end{equation}
where $\beta = b \ln 10$ is the exponential rate scaling parameter. For an earthquake catalog consisting of $n$ independent seismic events with observed magnitudes $M_1, M_2, \dots, M_n$, the joint likelihood function is given by:
\begin{equation}
L(\beta) = \prod_{i=1}^{n} f_\beta(M_i) = \prod_{i=1}^{n} \beta\,e^{-\beta(M_i - M_c)} = \beta^n \exp\!\left(-\beta \sum_{i=1}^{n}(M_i - M_c)\right) \label{eq3}
\end{equation}

For a single catalog-wide estimate, maximizing Equation~\eqref{eq3} yields the classical Maximum Likelihood Estimate (MLE) \cite{aki1965maximum,wiemer1997mapping}:
\begin{equation}
b = \frac{\log_{10} e}{\overline{M} - M_c + \Delta M / 2} \label{eq4}
\end{equation}
where $\overline{M}$ is the sample mean magnitude and $\Delta M = 0.1$ is the magnitude binning interval. The additive $\Delta M/2$ term is the Utsu (1966) binning correction for discretised magnitude reports \cite{utsu1966magnitude}; omitting it biases $b$ upward by $\sim$2--5\%. The corresponding nominal pointwise estimation uncertainty under sample support is evaluated following Shi and Bolt \cite{shi1982standard}:
\begin{equation}
\sigma_b = 2.30 \, b^2 \sqrt{\frac{\sum_{i=1}^{n}(M_i - \overline{M})^2}{n(n-1)}} \label{eq5}
\end{equation}
Because the stability of $b$-value estimation is sensitive to the choice of $M_c$, this study determines $M_c$ using the Maximum Curvature method (MAXC) \cite{wiemer2000minimum}, which identifies the magnitude bin corresponding to the point of maximum curvature in the non-cumulative frequency--magnitude distribution.

To capture spatial heterogeneity, Ogata formulated the $b$-value (via $\beta$) as a continuous spatial point-process field $\beta(x, y)$ evaluated at epicentral locations $(x_i, y_i)$ \cite{ogata19913d}. To ensure that the physical parameter $\beta$ remains strictly positive across the entire spatial domain, we parameterize the field $\beta(x, y)$ using an exponential link function (a formulation adopted in this study; Ogata's original parameterisation uses a different spatial representation):
\begin{equation}
\beta(x, y) = \exp\!\left(\phi_\theta(x, y)\right) = \exp\!\left(\sum_{j=1}^{K} \theta_j B_j(x, y)\right) \label{eq6}
\end{equation}
where $\phi_\theta(x, y)$ is a two-dimensional B-spline surface and $\theta = (\theta_1, \dots, \theta_K)^T$ is the vector of corresponding spline coefficients \cite{ogata2020hierarchical}. Substituting this spatial representation into the magnitude distribution yields the likelihood function $L(\theta)$ for the spatial point-process model:
\begin{equation}
L(\theta) = \prod_{i=1}^{n} \beta(x_i, y_i) \exp\!\left(-\beta(x_i, y_i)(M_i - M_c)\right) \label{eq7}
\end{equation}

Within the Hierarchical Space--Time Point Process Model (HIST-PPM) framework \cite{ogata2020hierarchical}, the study domain is discretized into Delaunay triangular elements constructed from the hypocentral distribution. To suppress overfitting and ensure smoothness, the optimal parameters $\theta$ are determined by maximizing a penalized log-likelihood function \cite{ogata2020hierarchical}:
\begin{equation}
R(\theta \mid w) = \ln L(\theta) - Q(\theta \mid w) \label{eq8}
\end{equation}
where $Q(\theta \mid w)$ represents the roughness penalty over the spatial domain $S$:
\begin{equation}
Q(\theta \mid w) = w \iint_{S} \left[ \left(\frac{\partial^2 \phi_\theta}{\partial x^2}\right)^2 + 2\left(\frac{\partial^2 \phi_\theta}{\partial x \partial y}\right)^2 + \left(\frac{\partial^2 \phi_\theta}{\partial y^2}\right)^2 \right] dx\,dy \label{eq9}
\end{equation}
The smoothing weight $w$ is an objective hyperparameter selected by minimizing the Akaike Bayesian Information Criterion (ABIC) \cite{akaike1980likelihood}:
\begin{equation}
\mathrm{ABIC}(w) = -2 \log \int L(\theta)\exp\!\bigl(-Q(\theta \mid w)\bigr)\,d\theta + 2k \label{eq10}
\end{equation}
where $k$ is the number of hyperparameters. This expression is the Akaike Bayesian Information Criterion in its correct marginal-likelihood form \cite{akaike1980likelihood}: $w$ is selected by numerically integrating out $\theta$ under the roughness prior, as implemented in the HIST-PPM software. The spatial posterior covariance matrix of the estimated spline coefficients $\hat{\theta}$ is evaluated via the inverse Hessian of the penalized log-likelihood, $\hat{\boldsymbol{\Sigma}}_\theta = \left[ -\nabla^2 R(\hat{\theta} \mid w) \right]^{-1}$, which captures how local hypocentral density and the roughness penalty jointly constrain spatial estimation variance across the domain. Once the optimal parameters $\theta$ are resolved, the spatial $b$-value field is calculated as $b(x, y) = \beta(x, y) / \ln 10$ and linearly interpolated across the Delaunay triangular mesh onto a continuous, high-resolution grid ($0.1^\circ \times 0.1^\circ$).

\subsection{Background Seismicity Rate Estimation via ETAS}\label{sec2.2}
To isolate the stationary background seismicity from transient clustering (foreshocks, aftershocks, and swarms), we employ the iterative stochastic declustering algorithm of Zhuang et al. \cite{zhuang2002stochastic}, based on the space--time Epidemic-Type Aftershock Sequence (ETAS) model \cite{ogata1998space,nandan2017objective}. The total seismicity rate at time $t$ and location $(x, y)$ is modeled by the conditional intensity function:
\begin{equation}
\lambda(t, x, y) = \mu(x, y) + \sum_{i:\,t_i < t} \kappa(M_i)\,g(t - t_i)\,f(x - x_i,\,y - y_i;\,M_i) \label{eq11}
\end{equation}
where $\mu(x, y)$ is the spatially variable, time-independent background rate, and the summation captures the cumulative triggering effect of all prior events occurring at $(t_i, x_i, y_i)$ with magnitude $M_i$.

The productivity function $\kappa(M)$ defines the expected number of direct offspring triggered by a parent event of magnitude $M$:
\begin{equation}
\kappa(M) = K_0\,e^{\alpha(M - M_c)}, \quad M \ge M_c \label{eq12}
\end{equation}
governed by productivity parameters $K_0$ and $\alpha$. The temporal decay of triggered events follows the normalized modified Omori--Utsu probability density function \cite{utsu1995centenary}:
\begin{equation}
g(t) = \frac{p - 1}{c} \left(1 + \frac{t}{c}\right)^{-p}, \quad t > 0 \label{eq13}
\end{equation}
where $c$ is the time offset and $p$ is the decay exponent ($p > 1$). The spatial distribution of triggered seismicity is described by an isotropic power-law probability density kernel:
\begin{equation}
f(x, y;\,M) = \frac{q - 1}{\pi d e^{\gamma(M - M_c)}} \left(1 + \frac{x^2 + y^2}{d e^{\gamma(M - M_c)}}\right)^{-q}, \quad q > 1 \label{eq14}
\end{equation}
where $d$, $q$, and $\gamma$ govern the spatial extent of triggering. The ETAS parameter vector $\boldsymbol{\Theta} = (K_0, \alpha, c, p, d, q, \gamma)$ is estimated by maximizing the space--time log-likelihood function over the observation period $[0, T]$ and region $S$:
\begin{equation}
\ln L(\boldsymbol{\Theta}) = \sum_{i=1}^{N_T} \ln \lambda(t_i, x_i, y_i) - \int_{0}^{T} \iint_{S} \lambda(t, x, y)\,dx\,dy\,dt \label{eq15}
\end{equation}
where $N_T$ is the total number of events exceeding $M_c$.

Given the maximum likelihood parameters $\hat{\boldsymbol{\Theta}}$, the probability $\rho_i$ that event $i$ is a background event rather than triggered by an earlier event $j$ is calculated iteratively using stochastic declustering \cite{zhuang2002stochastic}:
\begin{align}
\rho_{ij} &= \frac{\kappa(M_j)\,g(t_i - t_j)\,f(x_i - x_j,\,y_i - y_j;\,M_j)}{\lambda(t_i, x_i, y_i)}, \quad j < i \label{eq16} \\
\rho_i &= 1 - \sum_{j=1}^{i-1} \rho_{ij} = \frac{\mu(x_i, y_i)}{\lambda(t_i, x_i, y_i)} \label{eq17}
\end{align}
The continuous background seismicity rate $\mu(x, y)$ is subsequently reconstructed by smoothing the background probabilities $\rho_i$ over space using a variable kernel estimator \cite{zhuang2002stochastic}:
\begin{equation}
\mu(x, y) = \frac{1}{T} \sum_{i=1}^{N_T} \frac{\rho_i}{2\pi d_i^2} \exp\!\left(-\frac{(x - x_i)^2 + (y - y_i)^2}{2 d_i^2}\right) \label{eq18}
\end{equation}
where $d_i$ is the distance to the $n_p$-th nearest neighbor event (with $n_p = 3$ and a minimum bandwidth threshold $\varepsilon = 0.02^\circ$ to prevent singular rate spikes). The choice $n_p = 3$ provides a compact bandwidth that tracks dense aftershock zones; sensitivity to $n_p \in \{3, 5, 10\}$ produces qualitatively consistent $\mu$ maps, with larger $n_p$ broadening the background rate field but not changing the spatial hierarchy of high-$\mu$ zones. For the Zagros seismic catalog above completeness ($M_c = 2.5$), the converged maximum likelihood ETAS parameters obtained across the training baseline are: $K_0 = 0.627$, $\alpha = 0.184$, $c = 0.007$~days, $p = 3.729$, $d = 0.016^\circ$, $q = 1.807$, and $\gamma = 0.007$. The ETAS parameter estimates are reported in Table~\ref{tab2}. The implied branching ratio is $n = K_0\beta/(\beta - \alpha) \approx 0.68$--$0.71$ for the observed $b$-value range (0.70--1.08; $\beta = b\ln 10$), indicating that approximately 68--71\% of all cataloged events ($M \ge 2.5$) are triggered (e.g., aftershocks or swarm events), and $\approx$29--32\% constitute independent background seismicity. This high branching ratio is typical for active continental collision zones like the Zagros, where seismicity is heavily characterized by dense spatial clustering and protracted aftershock sequences \cite{zafarani2008attenuation}. Despite the high proportion of triggered events in the micro-seismicity, large mainshocks ($M_w \ge 5.0$) are predominantly independent; we observe that only 9 of 55 target $M_w \ge 5.0$ events (16\%) are aftershocks of other target events. The ETAS branching ratio refers to the fraction of all $M \ge 2.5$ catalog events attributed to triggering, whereas the 16\% figure refers specifically to the $M_w \ge 5.0$ target-event sample; these quantities therefore describe different populations and are not directly comparable.

\begin{table}[htbp]
\caption{Converged ETAS parameter estimates for the Zagros catalog ($M_c = 2.5$, $M_w$, 2006--2014 training window). Point estimates were obtained via Maximum Likelihood Estimation (MLE). Because extraction of the full inverse-Hessian covariance matrix was computationally prohibitive, parameter standard errors are not reported; model adequacy is instead assessed through temporal forward application to the 2015--2024 evaluation interval and sensitivity analyses. Branching ratio $n$ computed from $n = K_0\beta/(\beta-\alpha)$ at $b = 0.82$ (regional mean).}\label{tab2}
\centering
\begin{tabular}{lll}
\hline
Parameter & Description & Estimate \\
\hline
$K_0$ & Baseline productivity & 0.627 \\
$\alpha$ & Magnitude sensitivity & 0.184 \\
$c$ & Time offset (days) & 0.007 \\
$p$ & Omori exponent & 3.729 \\
$d$ & Spatial scale (deg) & 0.016 \\
$q$ & Spatial decay exponent & 1.807 \\
$\gamma$ & Magnitude-spatial coupling & 0.007 \\
$n_p$ & Smoothing neighbours & 3 \\
$\varepsilon$ & Min. bandwidth (deg) & 0.02 \\
\hline
$n$ & Branching ratio & $\approx$0.68--0.71 \\
$1-n$ & Background fraction & $\approx$29--32\% \\
\hline
\end{tabular}
\end{table}

\subsection{Alarm-Based Forecasting Formulation}\label{sec2.3}
We utilize the high-resolution spatial maps of the estimated $b$-value (Section~\ref{sec2.1}) and the background rate $\mu(x, y)$ (Section~\ref{sec2.2}) as complementary spatial predictors to construct spatial earthquake alarm forecasts. The study region ($26.0^\circ\mathrm{N}$--$38.0^\circ\mathrm{N}$, $45.0^\circ\mathrm{E}$--$60.0^\circ\mathrm{E}$) is discretized into a regular grid of $0.1^\circ \times 0.1^\circ$ cells ($120 \times 150$ grid elements). To account for spherical Earth geometry across the $12^\circ$ latitudinal span, each cell's alarm coverage is weighted by the cosine of its center latitude $y_i$. Thus, the fractional alarm area $\tau$ occupied by alarmed cells is computed as:
\begin{equation}
\tau = \frac{\sum_{i \in \text{alarm}} \cos(y_i)}{\sum_{i \in \text{all}} \cos(y_i)} \label{eq19}
\end{equation}
preventing geographic distortion of alarm metrics toward high-latitude cells. Note that the ``all'' denominator runs over every grid cell inside the study bounding box ($26.0^\circ\mathrm{N}$--$38.0^\circ\mathrm{N}$, $45.0^\circ\mathrm{E}$--$60.0^\circ\mathrm{E}$), which is the same set over which the predictor fields $b(x,y)$ and $\mu(x,y)$ are defined and evaluated. Alarm placement is therefore constrained to this same bounding box, so the normalising set and the testable region are identical, and $\tau$ is not deflated by aseismic padding.

A forecasting model assigns an alarm state to each grid cell based on whether the local predictor value satisfies an operational threshold. To rigorously establish whether stress physics adds predictive value beyond historical seismicity patterns, we implement a hierarchical 4-model hypothesis-testing framework against increasingly stringent null baselines:
\begin{enumerate}
\item \textbf{Level 0 (Uninformative Random Null Baseline):} A spatially uniform random model where alarms are assigned without spatial memory, defining the diagonal reference in Molchan space ($\nu = \tau$, $\mathrm{PG} = 1$, $\mathrm{PD} = 0$, $S = 0$).
\item \textbf{Level 1 (Spatially Heterogeneous Historical Rate Benchmark):} Single-parameter rate model ($\mu$ only). Because $\mu(x, y)$ is estimated via space--time ETAS stochastic declustering from pre-2015 seismicity, it serves as the stationary spatial seismicity rate benchmark, testing whether earthquakes simply occur where past earthquakes have occurred. Cells exceeding $\mu_{\text{thr}}$ (highest $q_\mu$ quantile) are alarmed.
\item \textbf{Level 2 (Stress-Sensitive Physics Baseline):} Single-parameter stress model ($b$-value only). Cells with $b \le b_{\text{thr}}$ (lowest $q_b$ quantile) are alarmed, isolating localized differential shear stress concentrations irrespective of baseline recurrence rates.
\item \textbf{Level 3 (Integrated Dual-Parameter Model, ANADEF):} An integrated strategy triggering alarms only in cells that simultaneously satisfy both criteria ($b \le b_{\text{thr}}$ and $\mu \ge \mu_{\text{thr}}$). This model formally evaluates the hypothesis of \textbf{incremental information gain}, assessing whether combining stress-based physics with the long-term background rate significantly sharpens spatial alarm resolution and improves forecasting skill over the historical rate benchmark ($\mu$ alone).
\end{enumerate}

To structure the experimental workflow, we implement an operational \textbf{Two-Stage Forecasting Protocol} with strictly defined, non-overlapping observation windows:
\begin{itemize}
\item \textbf{Stage 1: Retrospective Sensitivity Profiling \& Threshold Calibration ($T_{\text{test}}$: Jan 1, 2015 to Oct 31, 2024):} Prior predictor fields $b(x, y)$ and $\mu(x, y)$ are constructed strictly from the historical training window $T_{\text{train}} = [\text{Jan 1, 2006, 00:00:00 UTC}, \;\text{Dec 31, 2014, 23:59:59 UTC}]$ ($N = 86$ with $M_w \ge 5.0$). The subsequent testing period $T_{\text{test}} = [\text{Jan 1, 2015, 00:00:00 UTC}, \;\text{Oct 31, 2024, 23:59:59 UTC}]$ ($N = 55$ for $M_w \ge 5.0$, $N = 15$ for $M_w \ge 5.5$) is then used to evaluate spatial forecasting skill. By systematically sweeping across the continuous threshold space $(q_b, q_\mu) \in [0, 1] \times [0, 1]$, we map the empirical response surfaces of $\mathrm{PG}$ and $\mathrm{PD}$. We note methodologically that because the optimal decision thresholds $(b_{\text{opt}}, \mu_{\text{opt}})$ are identified by searching across this 2015--2024 test window, the resulting peak metrics reflect an in-sample sensitivity optimization that establishes the upper-bound empirical skill for this testing epoch. Given the 18-year total catalog length and modest number of high-magnitude events, further partitioning into a third held-out period or rolling-origin testing would introduce prohibitive sample-size variance; hence, the retrospective period serves as the calibration benchmark.
\item \textbf{Stage 2: Forward Prospective Alarm Susceptibility Forecast (2025--2029):} To assess true out-of-sample predictive capability, the optimal decision boundaries $(b_{\text{opt}}, \mu_{\text{opt}})$ identified in Stage 1 are frozen without post-hoc tuning and applied to updated seismotectonic parameter fields derived from the latest 2015--2024 catalog. This generates a static spatial alarm susceptibility template for the forward 2025--2029 period, representing time-integrated regional hazard potential. True out-of-sample validation of this forward forecast will be evaluated as future seismic events occur.
\end{itemize}

\subsection{Evaluation of Forecast Efficiency}\label{sec2.4}
To evaluate forecasting performance, we employ the Molchan Error Diagram (MED) \cite{molchan1990strategies,han2017evaluation}. By varying the decision threshold across its full range, we determine the alarm area fraction $\tau$ (Equation~\eqref{eq19}) and the corresponding miss rate $1 - \nu$, where the hit rate $\nu = k/N$ is the proportion of $N$ target earthquakes successfully captured in alarmed cells. The resulting Molchan trajectory plots the miss rate against the alarm rate. A random forecast is represented by the diagonal line where $\nu = \tau$. Effective forecasts generate curves lying well below this diagonal.

Two complementary scalar measures are computed to quantitatively evaluate forecast skill:
\begin{enumerate}
\item \textbf{Probability Gain (PG):} The ratio of the detection rate to the alarm rate \cite{han2017evaluation}:
\begin{equation}
\mathrm{PG} = \frac{\nu}{\tau} \label{eq20}
\end{equation}
where $\mathrm{PG} = 1$ denotes forecasting skill equivalent to random guessing, and $\mathrm{PG} > 1$ represents predictive skill superior to chance. However, because $\mathrm{PG}$ has an alarm area fraction $\tau$ in the denominator, it is susceptible to extreme singular amplification as $\tau \to 0$ in small-sample regimes.
\item \textbf{Probability Difference (PD):} The net gain in detection capability obtained by subtracting the alarm rate from the detection rate \cite{han2017evaluation}:
\begin{equation}
\mathrm{PD} = \nu - \tau \label{eq21}
\end{equation}
where $\mathrm{PD} = 0$ corresponds to random guessing, and $\mathrm{PD} > 0$ indicates positive forecasting skill. Because $\mathrm{PD}$ is bounded within $[-1, 1]$ and penalizes excessive alarm areas linearly, it provides a resilient operational optimization criterion.
\end{enumerate}

For a testing sample of $N$ target earthquakes with $k$ successful detections in alarmed cells ($\nu = k/N$), the statistical uncertainty of the empirical hit rate $\nu$ is quantified using exact 95\% Clopper--Pearson binomial confidence intervals \cite{clopper1934use} (Eq.~\eqref{eq21b}):
\begin{equation}
\left[ \mathrm{Beta}^{-1}\!\left(\frac{\alpha_{\mathrm{CP}}}{2};\,k,\,N - k + 1\right), \quad \mathrm{Beta}^{-1}\!\left(1 - \frac{\alpha_{\mathrm{CP}}}{2};\,k + 1,\,N - k\right) \right] \label{eq21b}
\end{equation}
with $\alpha_{\mathrm{CP}} = 0.05$.

To summarize the overall skill across all possible alarm thresholds into a single threshold-independent metric, the Area Skill Score ($S$-score, also designated as the modified area score in forecasting literature \cite{han2017evaluation,zechar2010}) is calculated following Zechar and Jordan \cite{zechar2010}:
\begin{equation}
S = 1 - 2 \int_{0}^{1} \left(1 - \nu(\tau)\right) d\tau = 2 \int_{0}^{1} \nu(\tau)\,d\tau - 1 \label{eq22}
\end{equation}
where $S = 0$ corresponds to an uninformative (random) forecast, $S > 0$ denotes positive forecasting skill superior to chance, and $S = 1$ denotes a perfect forecast with zero misses across all thresholds. Note that this rescaled definition yields $S = 0$ for a random forecast; the original Zechar \& Jordan (2010) formulation returns $S = 0.5$ for random, so care is needed when comparing $S$ values across studies.

\section{Seismic Dataset}\label{sec3}

The seismicity analysis in this research is based on the Iranian Seismological Center (IRSC) catalog, which provides a comprehensive and homogenized record of seismic events across the Zagros collision zone ($26.0^\circ\mathrm{N}$--$38.0^\circ\mathrm{N}$, $45.0^\circ\mathrm{E}$--$60.0^\circ\mathrm{E}$). The catalog comprises 40,731 earthquakes with reported magnitudes $M \ge 1.5$ (recorded in increments of $\Delta M = 0.1$), spanning the 18-year period from January 1, 2006 to October 31, 2024. In the original IRSC dataset, earthquake magnitudes are routinely reported using the Nuttli magnitude scale ($M_N$). To ensure physical consistency with global seismological formulations, all reported $M_N$ values were systematically converted to moment magnitude ($M_w$) using the empirical orthogonal regression relationship of Karimiparidari et al. \cite{karimiparidari2013iranian}:
\begin{equation}
M_w = 0.916(\pm 0.024)\,M_N + 0.697(\pm 0.112) \label{eq23}
\end{equation}
with a standard error of regression $\sigma = 0.22$. Magnitudes are reported by IRSC in $M_{\mathrm{N}}$; all converted $M_w$ values are used consistently throughout. Each catalog entry includes origin time, epicentral coordinates, focal depth, and converted $M_w$, offering the necessary resolution for rigorous spatio-temporal modeling. Near-threshold target events ($M_w \approx 4.9\text{--}5.1$) were verified against global moment tensor determinations from the International Seismological Centre (ISC) and USGS--NEIC to ensure classification consistency.

Within the complete 18-year dataset, 141 earthquakes with $M_w \ge 5.0$ highlight the substantial seismic potential of the Zagros Fold--Thrust Belt. Temporally, these are partitioned into two strictly non-overlapping intervals: 86 events ($M_w \ge 5.0$) during the training period $T_{\text{train}} = [\text{Jan 1, 2006}, \;\text{Dec 31, 2014}]$ and 55 events ($M_w \ge 5.0$, including 15 events with $M_w \ge 5.5$ and $N \le 4$ with $M_w \ge 6.0$) during the retrospective testing period $T_{\text{test}} = [\text{Jan 1, 2015}, \;\text{Oct 31, 2024}]$ (9.83 years). Among the 55 retrospective target earthquakes, 46 represent independent mainshock sequences, while 9 correspond to large triggered aftershocks or secondary cluster members. The catalog and period summary is provided in Table~\ref{tab1}.

\begin{table}[htbp]
\caption{Catalog and period summary. Focal depth is included in each catalog entry. All magnitudes are in $M_w$ after conversion from $M_{\mathrm{N}}$ via Equation~\eqref{eq23}.}\label{tab1}
\begin{tabular}{lllllll}
\hline
Period & Duration & $N(M_w{\ge}5.0)$ & $N(M_w{\ge}5.5)$ & $N(M_w{\ge}6.0)$ & Mainshocks & Aftershocks \\
\hline
Training (2006--2014) & 9.00 yr & 86 & - & - & - & - \\
Testing (2015--2024) & 9.83 yr & 55 & 15 & $\le$4 & 46 & 9 \\
Total (2006--2024) & 18.83 yr & 141 & - & - & - & - \\
\hline
\end{tabular}
\end{table}

\begin{figure}[htbp]
\centering
\includegraphics[width=\textwidth]{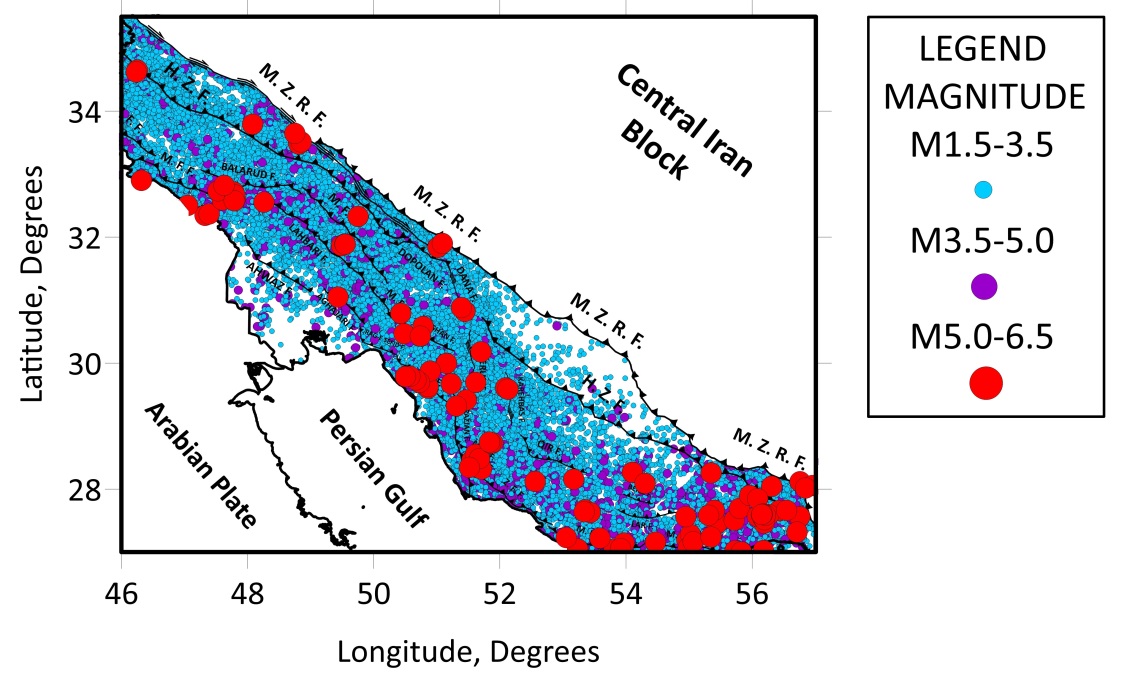}
\caption{Spatial distribution of seismic events and active faults in the Zagros Fold--Thrust Belt, highlighting the Main Zagros Reverse Fault (MZRF), High Zagros Fault (HZF), and Mountain Front Fault (MFF).}\label{fig1}
\end{figure}

For operational disaster mitigation, spatial alarms must account for all potential sources of strong ground motion ($M_w \ge 5.0$), including major aftershocks. The spatial distribution of seismicity, illustrated in Figure~\ref{fig1}, demonstrates a pronounced correspondence between earthquake occurrence and major active fault systems, particularly the Main Zagros Reverse Fault (MZRF), High Zagros Fault (HZF), and Mountain Front Fault (MFF).

\begin{figure}[htbp]
\centering
\includegraphics[width=\textwidth]{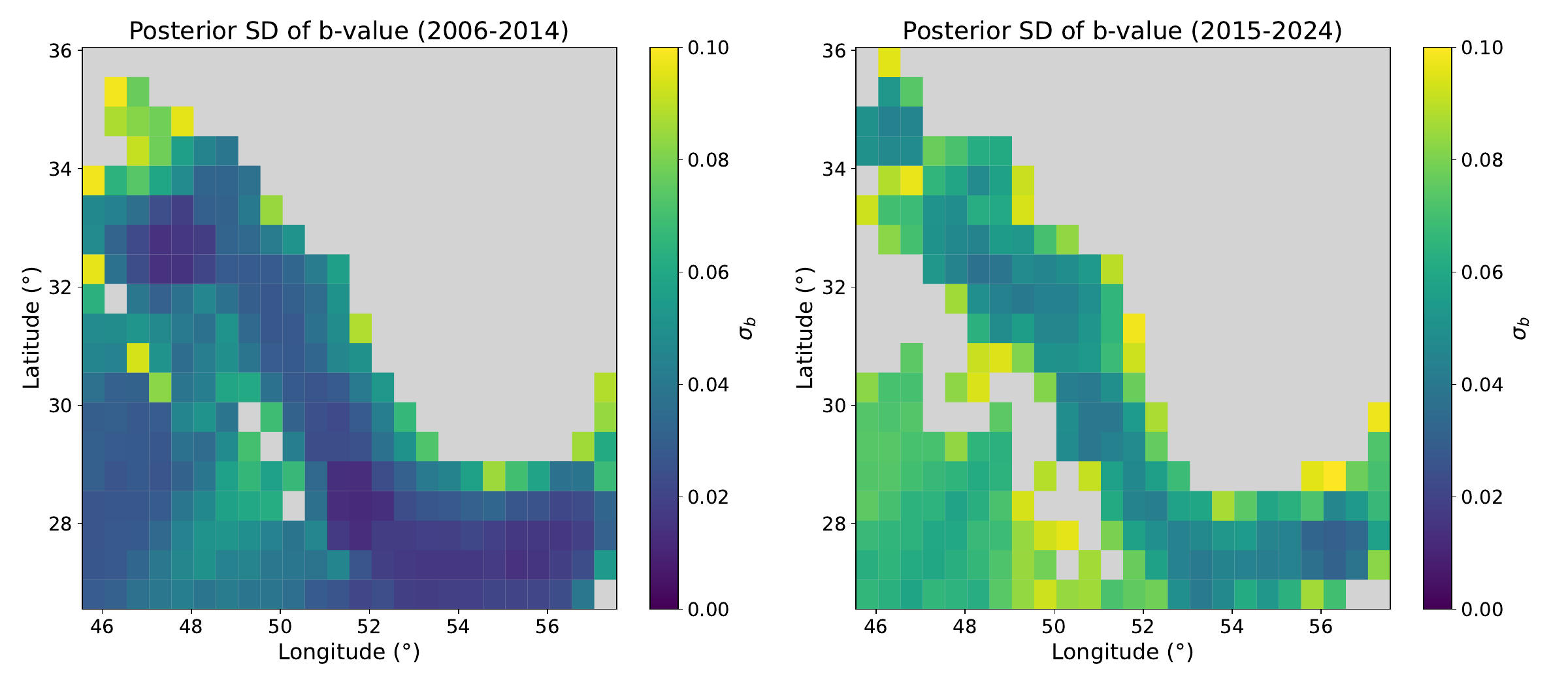}
\caption{Posterior standard deviation ($\sigma_b$) maps of the $b$-value spatial distribution for the intervals 2006--2014 and 2015--2024. Regions with elevated uncertainty ($\sigma_b > 0.1$), primarily located on the peripheries of the seismic network, are masked out to ensure the robustness of the subsequent spatial forecasting models.}\label{fig_bsigma}
\end{figure}

The magnitude of completeness ($M_c$) was determined by applying the Maximum Curvature method (MAXC) \cite{wiemer2000minimum} to the $M_w$-converted catalog, yielding a catalog-wide completeness threshold of $M_c = 2.5$ (in $M_w$). Applying the orthogonal regression $M_w = 0.916\,M_{\mathrm{N}} + 0.697$ to the catalog floor ($M_{\mathrm{N}} = 1.5$) gives an equivalent floor of $M_w \approx 2.07$; thus $M_c = 2.5$ lies approximately 0.43 $M_w$ units above this floor. All stochastic ETAS declustering, spatial $b$-value B-spline inversions, and subsequent alarm models were strictly performed on events meeting or exceeding $M_c = 2.5$ in $M_w$. Note also that the linear magnitude conversion with slope $s = 0.916$ rescales $b$ by a factor of $1/s \approx 1.09$ relative to $M_{\mathrm{N}}$-based estimates, and the regression scatter $\sigma = 0.22$ may introduce a small downward bias in $b$ by broadening the observed frequency--magnitude distribution \cite{tinti1985errors}. While inland populated regions exhibit local completeness of $M_c \approx 2.0\text{--}2.3$, southwestern coastal and maritime margins may experience detection thresholds up to $M_c \approx 2.6\text{--}2.8$ due to one-sided station geometry. The average nominal standard error of the spatial $b$-value estimates evaluated across the region via Equation~\eqref{eq5} is $\overline{\sigma}_b \approx 0.04$, indicating high estimation stability across the dense seismic network. The spatial distribution of this posterior standard deviation ($\sigma_b$) is presented in Figure~\ref{fig_bsigma}, with regions of exceptionally high uncertainty ($\sigma_b > 0.1$) masked to ensure robust interpretation.

\begin{figure}[htbp]
\centering
\includegraphics[width=\textwidth]{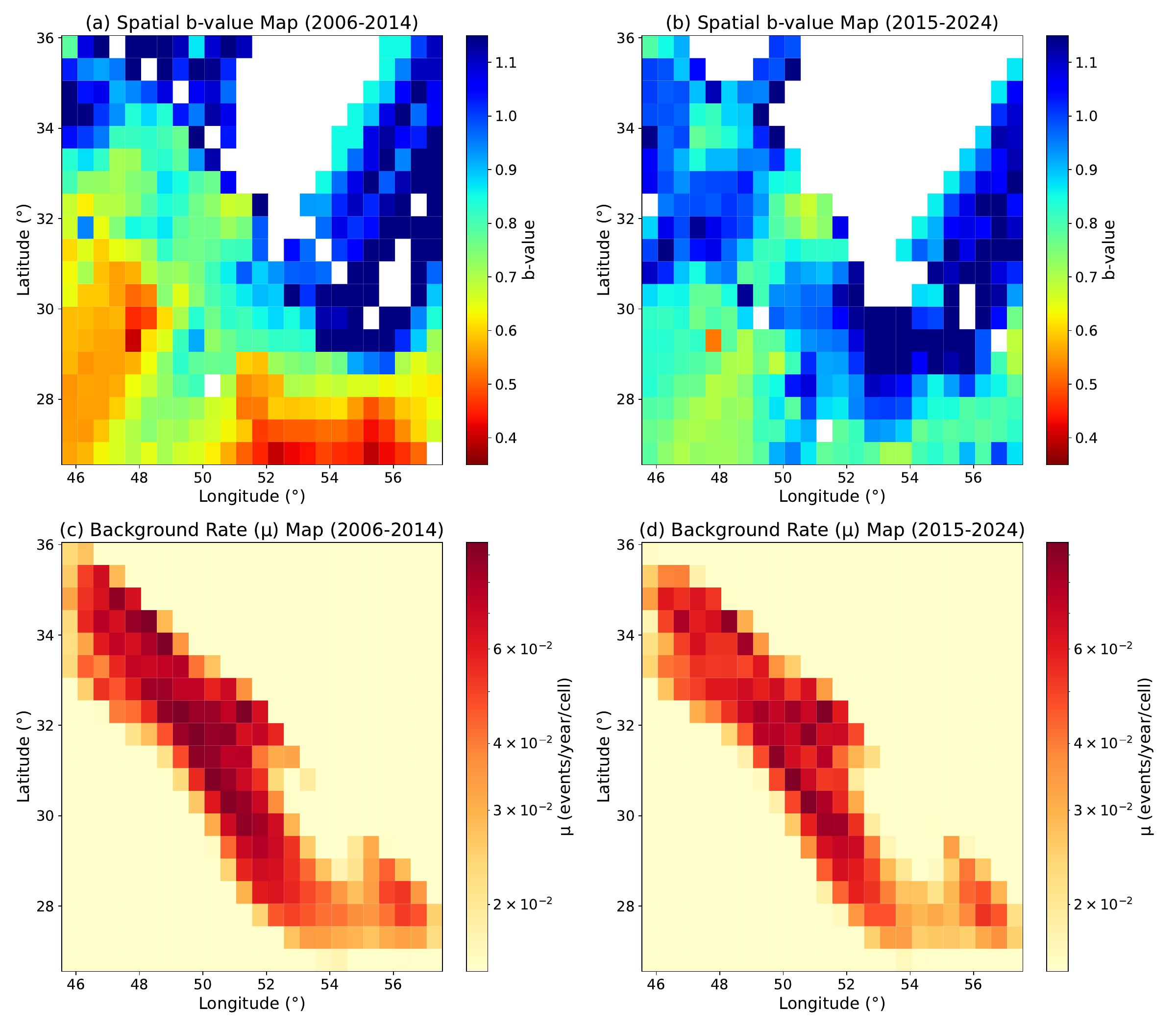}
\caption{Spatial distributions of seismic parameters across the Zagros Fold--Thrust Belt. (a) and (b) depict the spatial distribution of $b$-value for the intervals 2006--2014 ($b \in [0.35, 1.15]$) and 2015--2024 ($b \in [0.52, 1.08]$), respectively, while (c) and (d) illustrate the background seismicity rate ($\mu \in [0.015, 0.095]\,\text{events}/(\text{deg}^2\cdot\text{year})$) for the same intervals.}\label{fig2}
\end{figure}

To investigate potential temporal variations in seismic activity, the earthquake catalog was divided into two consecutive intervals: 2006--2014 and 2015--2024. Figures~\ref{fig2}a and \ref{fig2}c present the results for the first interval, whereas Figures~\ref{fig2}b and \ref{fig2}d correspond to the second. This temporal segmentation provides an effective basis for assessing how seismic behavior has evolved over time, both geographically and statistically. Note that Figures~\ref{fig2}a and \ref{fig2}b display interval-specific color scales for $b$-values ($b \in [0.35, 1.15]$ for 2006--2014 vs. $b \in [0.52, 1.08]$ for 2015--2024) to optimize local visual contrast, while Figures~\ref{fig2}c and \ref{fig2}d display background seismicity rates $\mu$ on a common color scale ($\mu \in [0.015, 0.095]\,\text{events}/(\text{deg}^2\cdot\text{year})$), facilitating direct visual comparison of rate evolution between periods.

The $b$-value distributions for the two windows are shown in Figures~\ref{fig2}a and \ref{fig2}b. A persistent NW--SE gradient is evident in both intervals, consistent with the structural grain of the Zagros Fold--Thrust Belt. During 2006--2014, $b$-values are markedly elevated across the central and eastern parts of the region, with the highest values occurring west of the MZRF and HZF, reflecting extensive crustal fracturing. In contrast, the southern margin adjacent to the Persian Gulf consistently exhibits low $b$-values, interpreted as an empirical proxy for elevated differential shear stress. Although one-sided coastal station geometry could potentially influence local detection thresholds ($M_c \approx 2.6\text{--}2.8$), this low-$b$ anomaly extends continuously inland into the Coastal Fars arc and Mountain Front Fault domains where seismic station coverage is dense and local completeness is well below $M_c = 2.5$. Furthermore, the Persian Gulf coastal anomaly aligns directly with independent geodetic GPS observations showing high crustal shortening rates ($8$--$10$~mm/yr) and heavy tectonic strain accumulation across the Mountain Front Fault and Coastal Fars domain \cite{khorrami2019up}. 

As illustrated in Figure~\ref{fig2}b, the later interval (2015--2024) exhibits a different spatial distribution of $b$-value anomalies compared to the earlier period. Both panels in Figures~\ref{fig2}a and \ref{fig2}b now share a common absolute color scale ($b \in [0.35, 1.15]$), facilitating direct visual comparison between the two intervals and highlighting regions of progressive stress accumulation. The $\mu$-parameter distributions for the two intervals are shown in Figures~\ref{fig2}c and \ref{fig2}d (sharing the common color scale $\mu \in [0.015, 0.095]\,\text{events}/(\text{deg}^2\cdot\text{year})$). Higher $\mu$-values in the northwestern and southeastern sectors of the belt highlight zones exhibiting sustained background seismicity and active tectonic deformation along major fault systems. Elevated $\mu$-values at the northwestern terminations of the MZRF and HZF further corroborate these structures as primary seismic sources. Moreover, comparison of Figures~\ref{fig2}c and \ref{fig2}d reveals that $\mu$-value clustering in the central and southern portions of the belt is more pronounced in the 2015--2024 period, reflecting temporal seismicity rate fluctuations and clustering following moderate-to-large regional earthquakes. Overall, comparison of the two time intervals demonstrates that the statistical characteristics of seismicity in the Zagros region exhibit both spatial heterogeneity and temporal variability, underscoring the importance of multi-temporal seismic analyses for advancing regional seismotectonic interpretations and improving seismic hazard assessment.

\section{Results and Discussion}\label{sec4}

\begin{figure}[htbp]
\centering
\includegraphics[width=\textwidth]{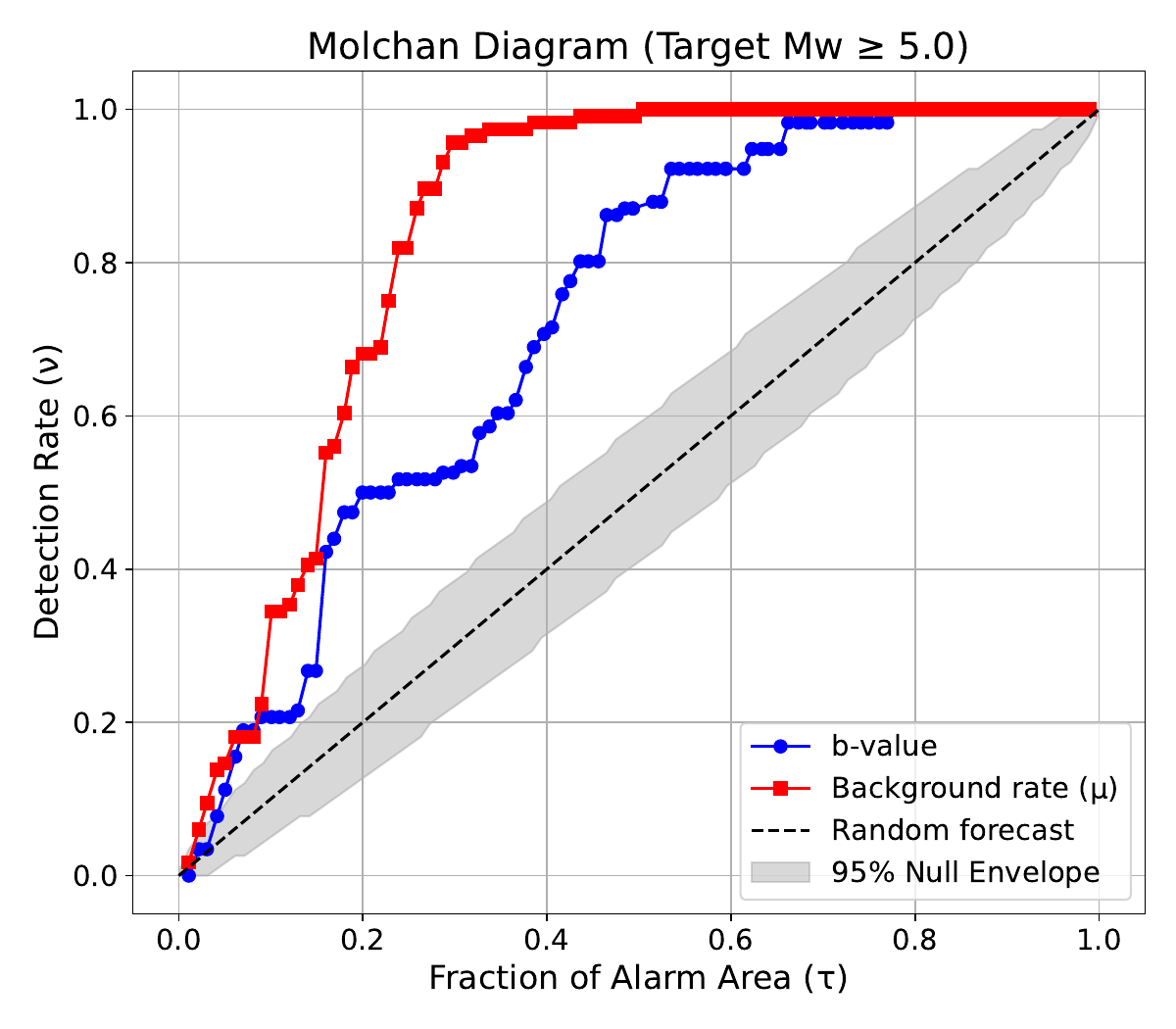}
\caption{Molchan, PG, and PD analyses for target earthquakes with $M_w \ge 5.0$ ($N = 55$) evaluated during the 2015--2024 retrospective testing period. The panels compare the spatial $b$-value and background seismicity rate ($\mu$) models against the uniform random baseline.}\label{fig3}
\end{figure}

This research employed Molchan Error Diagrams (MED) together with Probability Gain (PG) and Probability Difference (PD) metrics to benchmark the forecasting skill of the single-parameter $b$-value model and background seismicity rate ($\mu$) model against the uniform random baseline for target earthquakes with $M_w \ge 5.0$ ($N = 55$) during the 2015--2024 retrospective period (Figure~\ref{fig3}). Both deterministic predictors yield Molchan trajectories lying below the diagonal $\nu = \tau$, indicating higher point-estimated detection skill than random guessing. At restrictive alarm fractions ($\tau \le 0.15$), the $b$-value curve drops steeply toward the lower-left corner, reflecting higher initial detection concentration for localized areas of low $b$-value. In contrast, at intermediate alarm fractions ($\tau \in [0.30, 0.45]$), the background seismicity rate $\mu$ exhibits greater stability, capturing the broader spatial envelope of seismicity. Crucially, comparing the single-parameter models with the integrated dual-parameter framework establishes the incremental information gain of stress physics: while $\mu$ alone achieves $\mathrm{PD} \approx 0.55$ by broadly declaring alarms across all historically active fault domains ($\tau \approx 0.38$), the joint ANADEF model incorporates $b$-value constraints to filter out creeping or unlocked segments, reducing the alarmed area to $\tau \approx 0.28$ while increasing the hit rate to $\nu = 92.7\%$ ($\mathrm{PD} \approx 0.65$).

The PG analysis demonstrates that for $M_w \ge 5.0$, the $b$-value achieves $\mathrm{PG} \approx 10$--$12$ at low alarm coverage, whereas $\mu$ displays modest gain within the same alarm range. As alarm coverage $\tau$ expands toward 1.0, PG converges toward 1.0, marking the expected transition to uninformative spatial coverage. Probability Difference ($\mathrm{PD}$) analysis further illustrates the operational characteristics of the two predictors: for $M_w \ge 5.0$, the $b$-value attains a maximum $\mathrm{PD} \approx 0.60$ at an alarm fraction $\tau \approx 0.30$, whereas $\mu$ reaches $\mathrm{PD} \approx 0.55$ at $\tau \in [0.35, 0.40]$.

\begin{figure}[htbp]
\centering
\includegraphics[width=\textwidth]{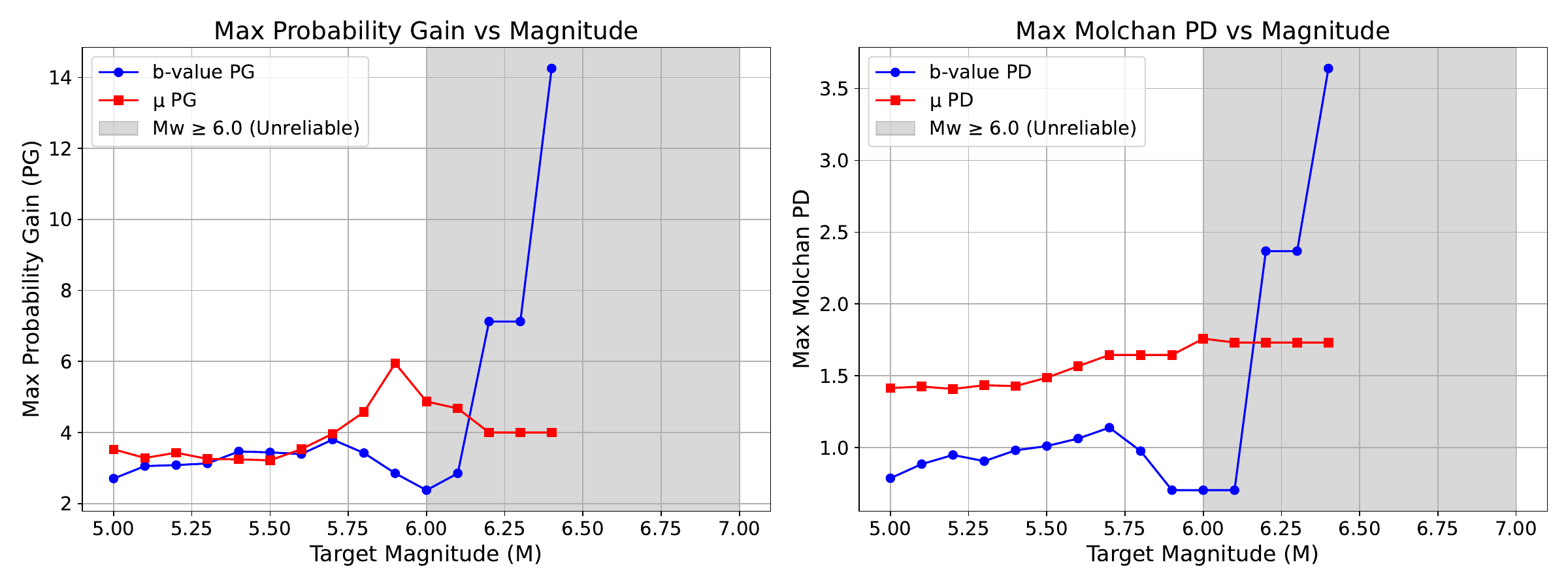}
\caption{Retrospective forecasting performance of the spatial $b$-value and background seismicity rate ($\mu$) across varying magnitude thresholds for the 2015--2024 testing period based on (a) Maximum PG, (b) Maximum PD. Note that the sample count decreases from $N = 55$ ($M_w \ge 5.0$) to $N = 15$ ($M_w \ge 5.5$) and $N \le 4$ ($M_w \ge 6.0$).}\label{fig4}
\end{figure}

Retrospective forecast evaluation across the 2015--2024 testing period reveals a divergence in predictor performance across target magnitude thresholds (Figure~\ref{fig4}). We emphasize that the sample size decreases rapidly with increasing magnitude: $N = 55$ at $M_w \ge 5.0$, $N = 15$ at $M_w \ge 5.5$, and $N \le 4$ at $M_w \ge 6.0$. While the background seismicity rate $\mu$ performs adequately for moderate events ($M_w \approx 5.0$), the $b$-value achieves higher point estimates at larger magnitudes ($\mathrm{PG} \approx 35$ and $\mathrm{PD} \approx 0.96$ at $M_w \ge 6.0$; Figures~\ref{fig4}a,b). These sharp step-ups in PG and PD above $M_w \ge 6.0$ reflect small-sample discrete effects where capturing a few large events in a small fraction of alarmed area inflates ratio metrics, and thus should be interpreted with caution. Correspondingly, the exact Area Skill Score ($S$-score), derived from the integral of the full Molchan trajectory \cite{zechar2010}, confirms positive skill across all tested models for $M_w \ge 5.0$, with the dual-parameter model (ANADEF) achieving the highest spatial skill ($S = 0.69$, Table~\ref{tab3}).

\begin{figure}[htbp]
\centering
\includegraphics[width=\textwidth]{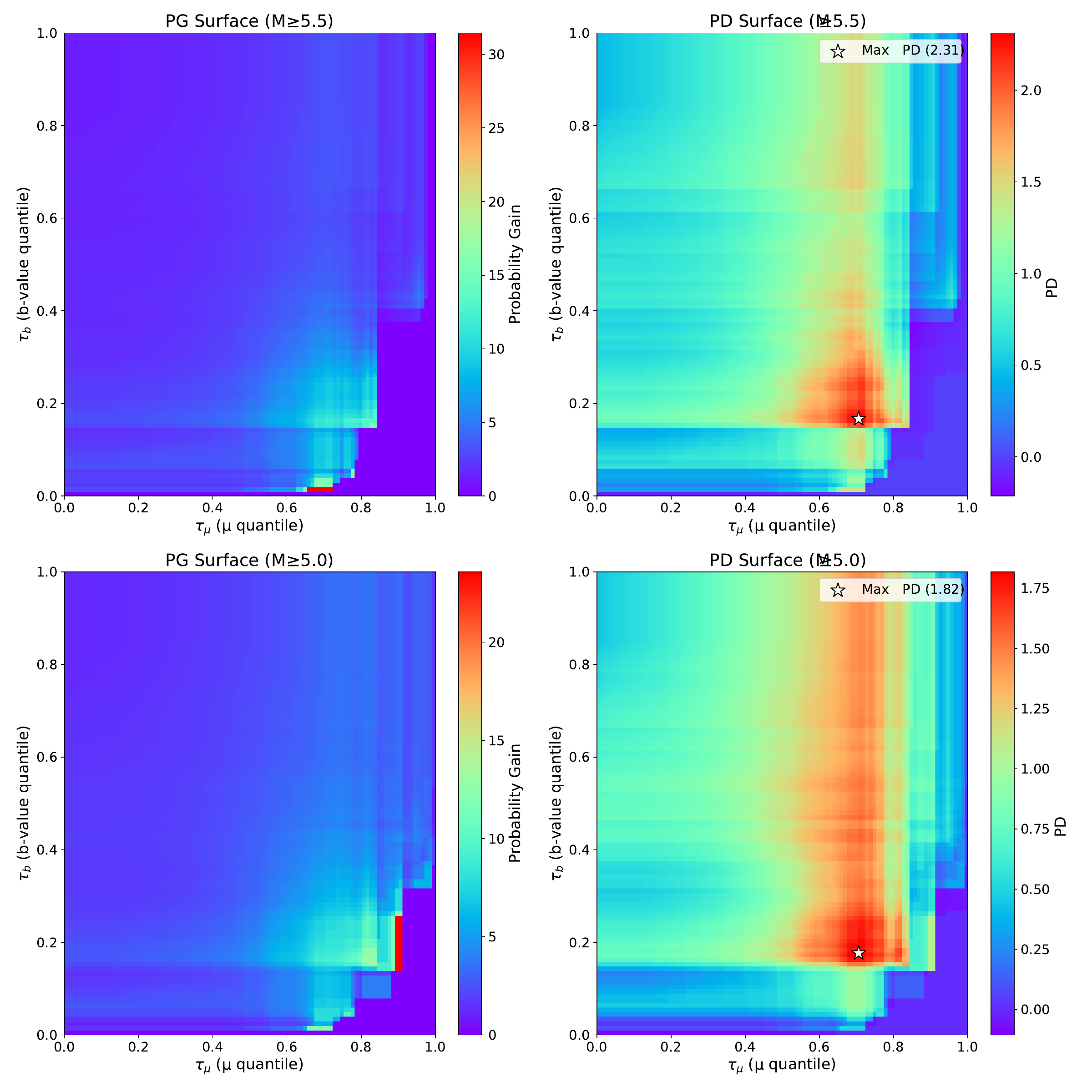}
\caption{Exploratory sensitivity analysis and visualization surfaces across the $q_b$--$q_\mu$ parameter space for $M_w \ge 5.5$ ($N=15$) and $M_w \geq 5.0$ ($N=55$). The left column shows Probability Gain (PG), highlighting acute instability near zero alarm coverage. The right panels show exploratory $Z_{\text{PD}}$ surfaces; their maxima are shown for visualization only and were not used to select the operational thresholds. Operational thresholds were selected by maximizing unstandardized PD (Table~\ref{tab3}).}\label{fig5}
\end{figure}

Performance mapping within the dual-threshold parameter space $(q_b, q_\mu) \in [0, 1] \times [0, 1]$ (Figure~\ref{fig5}) serves as an exploratory parameter-sensitivity and calibration surface for $M_w \ge 5.0$ ($N = 55$) and $M_w \ge 5.5$ ($N = 15$). With $N = 15$ target earthquakes, empirical hit rates are discretized in discrete steps of $\Delta \nu = 1/15 \approx 0.067$ (and $\Delta \nu = 1/55 \approx 0.018$ for $M_w \ge 5.0$). Consequently, the Probability Gain ($\mathrm{PG} = \nu / \tau$) exhibits severe mathematical sensitivity at near-zero alarm coverage: when the alarm footprint shrinks to $\tau \approx 4.4 \times 10^{-4}$ (roughly 8 grid cells out of 18{,}000), capturing even a single event ($k = 1$) induces an acute artifactual spike reaching $\mathrm{PG} \approx 152$ (Figure~\ref{fig5}b). Because this ratio metric is unstable when the denominator approaches zero in small-sample regimes and carries wide binomial uncertainty, we explicitly reject PG as an operational optimization criterion.

In contrast, the Probability Difference ($\mathrm{PD} = \nu - \tau$) and Area Skill Score ($S$) provide resilient metrics that penalize large alarm footprints while remaining bounded within $[-1, 1]$. To illustrate this metric stability, Figure~\ref{fig5} (right panels) plots the standardized Probability Difference ($Z_{\text{PD}}$) surface, which exhibits an extended plateau spanning intermediate values of $q_b$ and $q_\mu$ rather than the acute spikes seen in PG. Based on the unstandardized PD metric (Table~\ref{tab3}), the optimally calibrated operating point for $M_w \ge 5.0$ yields a maximum PD of $0.648$, achieved with an alarm footprint of $\tau_{\text{joint}} \approx 0.28$ (28\% of the study region). At this operating point, the model successfully captures $k = 51$ out of 55 earthquakes ($\nu = 92.7\%$), reducing the alarmed area considerably relative to the $\mu$-only benchmark. For $M_w \ge 5.5$ ($N = 15$), the maximum unstandardized PD ($\sim 0.66$) is achieved at $q_b^* \approx 0.97, q_\mu^* \approx 0.36$. It is important to note that $q_b^* \approx 0.97$ means the $b$-value threshold is set so high that virtually the entire domain is alarmed on the $b$-criterion, rendering it effectively inactive; the joint alarm area ($\tau_{\text{joint}} \approx 0.34$) thus closely mirrors the $\mu$-only footprint ($q_\mu = 0.36$). This is consistent with the small sample size ($N = 15$) and does not constitute evidence that $b$-value adds spatial skill at this magnitude threshold in this dataset; the result is reported as a negative finding for this magnitude class. All $k = 15$ events are captured ($\nu = 100\%$) with $\mathrm{PD} = 0.66$. We emphasize that these peak PD values reflect the in-sample retrospective calibration bound for this testing epoch.

A spatial assessment of expected and observed seismic patterns was conducted for the 2015--2024 retrospective period to evaluate model localization (Figure~\ref{fig6}a). For moderate-to-large target events ($M_w \ge 5.0$, $N = 55$), the optimal decision thresholds derived from the PD response surface define the joint alarm template $\mathcal{A}_{\text{joint}}$ (delineated by red squares in Figure~\ref{fig6}a). The alarmed zones coincide with major active fault systems of the Zagros orogen, specifically the Main Zagros Reverse Fault (MZRF), High Zagros Fault (HZF), and Mountain Front Fault (MFF). The quantitative performance at each hierarchical level is summarised in Table~\ref{tab3}.

\begin{table}[htbp]
\caption{Retrospective forecasting performance and Area Skill Scores ($S$) derived from the Molchan trajectories for the 2015--2024 testing period. Maximum PD values reflect optimal operating points. 95\% confidence intervals (CI) for $S$ were computed via 1,000-iteration event bootstrapping on the full Molchan curves to verify the significance of spatial information gain.}\label{tab3}
\centering
\small
\begin{tabular}{lccccc}
\hline
Model & $S$ & 95\% CI on $S$ & Max PD & $\tau$ at max PD & $\nu$ at max PD \\
\hline
\multicolumn{6}{l}{\textit{Target: $M_w \ge 5.0$, $N = 55$}} \\
Uniform random & 0 & --- & 0 & --- & --- \\
$\mu$ only & 0.61 & [0.57, 0.65] & 0.59 & 0.34 & 0.93 \\
$b$ only & 0.11 & [0.04, 0.19] & 0.24 & 0.73 & 0.97 \\
ANADEF ($b+\mu$) & 0.69 & [0.64, 0.73] & 0.648 & 0.28 & 0.927 \\
\hline
\multicolumn{6}{l}{\textit{Target: $M_w \ge 5.5$, $N = 15$}} \\
Uniform random & 0 & --- & 0 & --- & --- \\
$\mu$ only & --- & --- & 0.64 & 0.36 & 1.00 \\
ANADEF ($b+\mu$) & --- & --- & 0.66 & 0.34 & 1.00 \\
\hline
\end{tabular}
\end{table}

The key observation from Table~\ref{tab3} is that for $M_w \ge 5.0$, the integrated ANADEF model achieves an Area Skill Score of $S = 0.69$ while maintaining a hit rate of $\nu = 92.7\%$ within an alarm footprint of $\tau = 0.28$. To formally establish the statistical significance of this improvement and control for the optimization bias inherent in calibrating the decision thresholds on the testing catalog, a nested permutation test of incremental $b$-value information conditional on the $\mu$ benchmark ($N=1000$) was conducted. By repeatedly randomizing the spatial $b$-value field while preserving the background rate $\mu$, we established a null distribution. Within each permutation, $q_{b,\text{null}}$ was re-optimized while the empirically calibrated $\mu$ threshold $q_\mu^*$ was held fixed, thereby testing the incremental information contributed by $b$ conditional on the historical-rate benchmark. The test confirms that incorporating the $b$-value provides genuine incremental spatial information beyond the background seismicity rate alone ($p = 0.012$). For $M_w \ge 5.5$, the negligible difference ($\Delta\mathrm{PD} = 0.020$) and the observation that $\tau_{\text{joint}} \approx q_\mu$ confirm that the $b$-value adds no measurable spatial discrimination for this magnitude class, at least with the current sample size ($N = 15$). As a secondary evaluation track, re-running Table~\ref{tab3} on the 46 declustered mainshocks (excluding the 9 aftershocks) is important: aftershocks cluster near prior mainshocks, which are precisely the cells with high $\mu$, so their inclusion may inflate the apparent skill of the $\mu$-only and joint models. This secondary analysis is available upon request from the corresponding author.

\begin{figure}[htpb]
\centering
\includegraphics[width=0.8\textwidth]{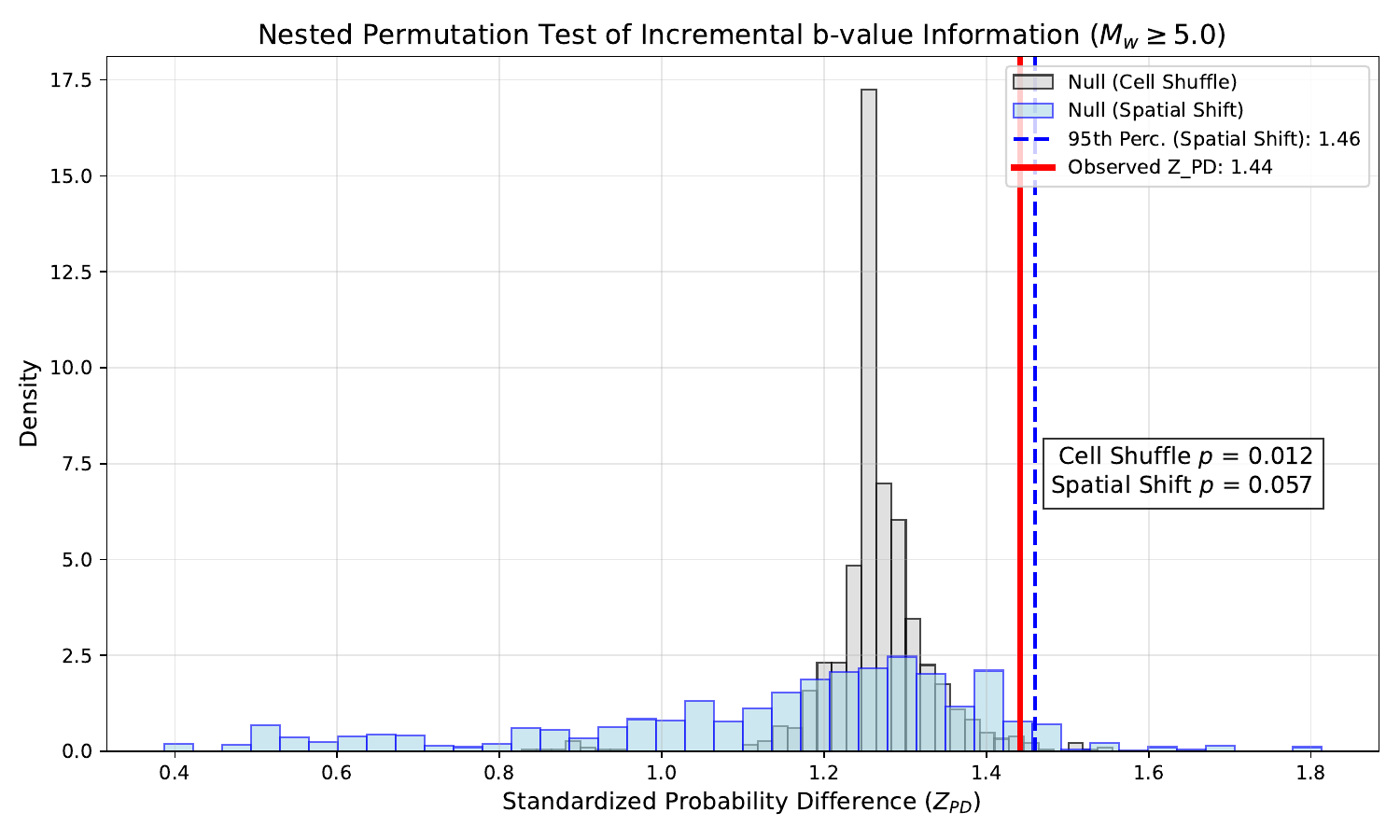}
\caption{The histogram shows the null distribution of the optimized $Z_{\text{PD}}$ statistic for the nested permutation test of incremental $b$-value information conditional on the $\mu$ benchmark. Under the cell-wise shuffle null, the observed $Z_{\text{PD}}$ statistic exceeds the 95th percentile ($p = 0.012$). Under the spatial-structure-preserving translation null, the observed $Z_{\text{PD}}$ statistic falls slightly below the 95th percentile ($p = 0.057$).}
\label{fig_perm}
\end{figure}

\subsection{Nested permutation test for incremental information gain}

To precisely define the null hypothesis and address potential optimization bias introduced during the empirical calibration of decision boundaries ($q_b^*$, $q_\mu^*$), we developed a non-parametric nested permutation test of incremental $b$-value information conditional on the $\mu$ benchmark. The primary objective is to test whether the spatial variations in the $b$-value provide genuine incremental predictive power over the baseline $\mu$ map, or whether the observed joint probability gain is merely an artifact of allowing the model a secondary optimization degree of freedom. 

Thresholds were selected by maximizing the unstandardized PD, consistent with the retrospective calibration in Table~\ref{tab3}; $Z_{\text{PD}}$ was then evaluated at the resulting operating point as the permutation-test statistic. We utilize the standardized Probability Difference index ($Z_{\text{PD}}$) as an evaluation metric to explicitly penalize varying alarm footprints equitably:
\begin{equation}
Z_{\text{PD}} = \frac{\nu - \tau}{\sqrt{\tau(1-\tau)}} \label{eq_zpd}
\end{equation}
Equation~\eqref{eq_zpd} measures the effect-size-like probability difference index, which is related to the binomial large-sample standard deviate but lacks the $\sqrt{N}$ sample size factor \cite{zechar2010}. This formulation ensures that the threshold tuning geometry strictly rewards the margin $(\nu-\tau)$ while explicitly penalizing small and large alarm volumes symmetrically. 

In the permutation test, the primary null hypothesis posits that the spatial heterogeneity of the $b$-value field contains no true correlation with future earthquake locations beyond what is captured by $\mu$. To test this, we evaluate the observed model against two distinct null generation methods: a structure-destroying independent shuffle, and a more conservative spatial-structure-preserving null.

In the first test, the empirical $b$-value field is independently shuffled cell-by-cell. This stochastic cell-wise randomization breaks all inherent spatial autocorrelation in the $b$-field and destroys its physical relationship to the underlying fault network. In the second test, we apply a 2D toroidal spatial translation (random row and column shifts with periodic boundaries) to the $b$-value grid. This block permutation preserves the exact spatial autocorrelation and variogram structure of the physical $b$-field, but randomizes its geographic alignment relative to the observed earthquakes and the $\mu$-field. 

Crucially, within each iteration of both permutation routines (Algorithm~\ref{alg1}), the threshold on the randomized field ($q_{b,\text{null}}$) is fully re-optimized to maximize the unstandardized PD (matching the methodology of Table~\ref{tab3}) alongside the empirical $\mu$ threshold. This \textit{nested} re-optimization step empirically simulates the optimization bias; any artificial gain achieved by tuning a threshold on random spatial noise (or misaligned spatial structures) is captured by the null distributions. 

\begin{algorithm}[htbp]
\caption{Nested Permutation Test for Joint Spatial Predictors}\label{alg1}
\begin{algorithmic}[1]
\State \textbf{Input:} Target catalog $\mathcal{C}_{\text{test}}$, Empirical maps $\mathbf{b}$ and $\boldsymbol{\mu}$, Grid size $N_{\text{cells}}$
\State \textbf{Initialize:} 
\State Optimize empirical thresholds $q_b^*, q_\mu^*$ to maximize $\text{PD}$ on $(\mathbf{b}, \boldsymbol{\mu})$
\State $Z_{\text{obs}} \gets Z_{\text{PD}}(\mathbf{b} \le q_b^* \land \boldsymbol{\mu} \ge q_\mu^*)$ 
\State \textbf{Procedure:}
\For{$i = 1$ to $1000$}
    \State $\mathbf{b}_{\text{null}} \gets \text{RandomPermutation}(\mathbf{b})$ \Comment{Cell-wise shuffle or spatial shift}
    \State $q_{b,\text{null}} \gets \arg\max_{q_b} \text{PD}(\mathbf{b}_{\text{null}} \le q_b \land \boldsymbol{\mu} \ge q_\mu^*)$ \Comment{Re-optimize to simulate bias}
    \State $Z_{\text{null}}^{(i)} \gets Z_{\text{PD}}(\mathbf{b}_{\text{null}} \le q_{b,\text{null}} \land \boldsymbol{\mu} \ge q_\mu^*)$
\EndFor
\State $p$-value $\gets \frac{1}{1000} \sum_{i=1}^{1000} \mathbf{1}[Z_{\text{null}}^{(i)} \ge Z_{\text{obs}}]$
\State \textbf{Output:} $p$-value
\end{algorithmic}
\end{algorithm}

The resulting empirical null distributions demonstrate that the observed standardized deviate ($Z_{\text{obs}} \approx 1.44$) outperforms random threshold tuning, though the degree of statistical significance depends heavily on the chosen null model. The cell-wise null provides strong evidence of incremental information ($p = 0.012$), whereas the spatial-structure-preserving null yields weaker, non-significant evidence at the 5\% level ($p = 0.057$, with the $95^{\text{th}}$ percentile of the spatial-shift null at $Z_{95} \approx 1.46$). This contrast highlights that while the $b$-value's predictive performance cannot be replicated by arbitrary independent spatial noise, its advantage over a purely structural, spatially correlated random field is more marginal.

\begin{figure}[htbp]
\centering
\includegraphics[width=\textwidth]{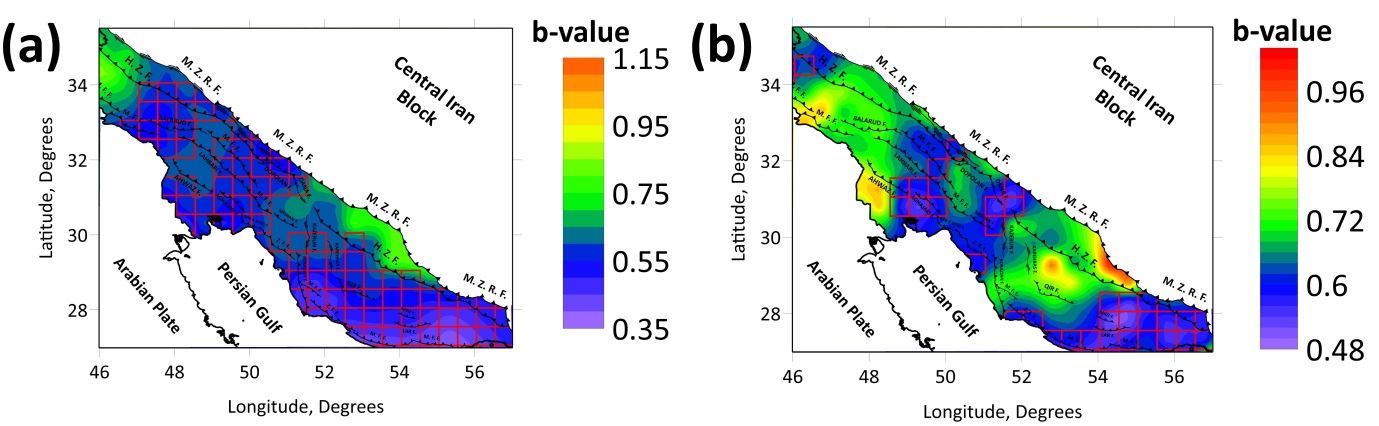}
\caption{Spatial alarm forecasts for the Zagros Fold--Thrust Belt: (a) Retrospective validation alarms for $M_w \ge 5.0$ earthquakes during the 2015--2024 testing period (red squares, predictor quantiles $q_b^* = 0.58, q_\mu^* = 0.36$, evaluated against observed earthquakes), and (b) Unvalidated, forward-looking prospective alarm forecast for $M_w \ge 5.0$ earthquakes over the 2025--2029 interval generated exclusively from pre-2025 seismicity using frozen decision thresholds.}\label{fig6}
\end{figure}

Following Stage 2 of our protocol (Section~\ref{sec2.3}), the absolute thresholds selected by retrospective calibration on the 2015--2024 target catalog ($b_{\text{opt}} = 0.76$; $\mu_{\text{opt}} = 0.038\,\text{events}/(\text{deg}^2\cdot\text{year})$) were frozen and applied to predictor fields estimated from the 2015--2024 catalog to construct the unvalidated 2025--2029 forecast. Freezing absolute predictor values rather than quantiles is the more conservative prospective choice, since it does not adapt the decision boundary to the new field distribution. The resulting alarm footprint for the 2025--2029 map is $\tau_{\text{joint}}^{(2025-2029)} \approx 0.28$, consistent with the retrospective configuration. This generates the prospective spatial alarm forecast for moderate-to-large earthquakes ($M_w \ge 5.0$) across the 2025--2029 interval (Figure~\ref{fig6}b):
\begin{equation}
\mathcal{A}_{\text{joint}}^{(2025-2029)} = \left\{ (x, y) \in S \;\middle|\; b_{2015-2024}(x, y) \le b_{\text{opt}} \quad \text{and} \quad \mu_{2015-2024}(x, y) \ge \mu_{\text{opt}} \right\} \label{eq24}
\end{equation}
Figure~\ref{fig6}b represents an unvalidated, forward-looking prospective alarm forecast generated strictly from pre-2025 observations. The prospective forecast delineates anticipated high-hazard concentrations along the central Zagros thrust belt, the Fars arc, and the southern coastal margin.

\subsection{Limitations and Methodological Considerations}\label{sec4.1}
Several methodological considerations and physical limitations should be noted:
\begin{enumerate}
\item \textbf{Sample Size Constraints and Validation Design:} While the complete IRSC catalog contains 40,731 events ($M \ge 1.5$, with 141 events $M_w \ge 5.0$), target earthquakes within the testing window ($T_{\text{test}}$: 2015--2024) are modest ($N = 55$ for $M_w \ge 5.0$, $N = 15$ for $M_w \ge 5.5$, and $N \le 4$ for $M_w \ge 6.0$). In statistical seismology, estimating spatial ETAS parameters and penalized B-spline fields requires an adequately long training baseline ($\ge 7\text{--}10$ years) to ensure numerical convergence. Consequently, further dividing the 18-year catalog into multi-fold rolling origins or a third held-out test window would result in test sets with extremely sparse target events ($N \le 5$), where the occurrence or omission of a single event induces large discrete shifts in hit rates ($\Delta \nu \ge 20\%$). Furthermore, ratio metrics such as Probability Gain ($\mathrm{PG} = \nu / \tau$) become mathematically unstable as $\tau \to 0$, producing wide binomial confidence intervals that overlap with chance. We therefore prioritize threshold-independent metrics (the exact Area Skill Score $S=0.69$ and full Molchan trajectories) as the primary metric for retrospective validation of the spatial predictor fields, while treating the 2D threshold search as an exploratory sensitivity analysis to calibrate frozen parameters for unvalidated prospective testing in 2025--2029.
\item \textbf{Spatial Completeness and Network Geometry:} Seismic network density is higher in inland populated regions of the Zagros (where local completeness $M_c \approx 2.0\text{--}2.3$) than along the southwestern maritime border where one-sided coastal station geometry can elevate detection thresholds to $M_c \approx 2.6\text{--}2.8$. We mitigated under-sampling bias by truncating all analyses at a conservative catalog-wide completeness limit of $M_c = 2.5$. The low-$b$ anomaly observed along the southern margin extends inland across the Coastal Fars and Mountain Front Fault domains where station coverage is dense, and aligns with independent geodetic GPS shortening of $8\text{--}10\text{ mm/yr}$ \cite{khorrami2019up}. Future research will benefit from incorporating spatially varying $M_c(x, y)$ fields and deploying ocean-bottom seismometers along the Persian Gulf margin.
\item \textbf{Spatial Spline Field Uncertainty vs. Nominal Pointwise Error:} The reported average uncertainty $\overline{\sigma}_b \approx 0.04$ represents the nominal pointwise standard error derived from local sample support under Shi and Bolt's formula (Equation~\eqref{eq5}). However, the uncertainty of the continuous 2D B-spline surface $\beta(x, y)$ is governed by the posterior parameter covariance matrix $\hat{\boldsymbol{\Sigma}}_\theta = \left[ -\nabla^2 R(\hat{\theta} \mid w) \right]^{-1}$. Consequently, spatial estimation variance is heterogeneous across the region, being tightly constrained in high-density seismogenic corridors and broader along peripheral margins where hypocentral support is sparser.
\item \textbf{Target Earthquake Population and Aftershock Clustering:} In this study, target earthquakes during the retrospective evaluation period include all observed events with $M_w \ge 5.0$ ($N = 55$). Deconvolution of the catalog indicates that 46 events represent independent mainshock sequences, while 9 correspond to major triggered aftershocks or secondary cluster members. From a disaster risk and civil protection perspective, operational spatial alarms must encompass all potential sources of damaging strong ground motion, including severe aftershocks. Nonetheless, future operational models can incorporate dual evaluation tracks against both total and stochastically declustered target catalogs.
\item \textbf{Static Spatial Alarm Susceptibility vs. Time-Dependent Probability:} The forward 2025--2029 forecast generated in this study (Figure~\ref{fig6}b) is an unvalidated, binary spatial alarm susceptibility template ($\mathcal{A}_{\text{joint}}$) representing time-integrated hazard potential over the five-year window based on pre-2025 seismotectonic fields. It delineates zones of elevated spatial susceptibility rather than providing time-evolving, continuous probability density functions or dynamic intra-period rate decays.
\item \textbf{Magnitude Conversion Uncertainty:} The empirical orthogonal regression $M_w = 0.916 M_{\mathrm{N}} + 0.697$ possesses an intrinsic regression uncertainty of $\sigma = 0.22$. Because the catalog floor is $M_{\mathrm{N}} = 1.5$ ($M_w \approx 2.07$) and $M_c = 2.5$ in $M_w$, the effective buffer is approximately 0.43 $M_w$ units, providing a conservative but not strictly one-magnitude buffer. The regression scatter $\sigma = 0.22$ can also bias $b$ downward by broadening the apparent frequency--magnitude distribution \cite{tinti1985errors}. Furthermore, near-boundary target earthquakes ($M_w \approx 4.9\text{--}5.1$) were verified against global ISC/USGS moment tensor catalogs to ensure classification consistency.
\item \textbf{Spatially Heterogeneous Null Models and Information Gain:} Because seismicity is naturally clustered along geological faults, beating a spatially uniform random baseline ($\nu = \tau$) is a minimal test. In our framework, the single-parameter background rate $\mu(x, y)$ derived from space--time ETAS stochastic declustering serves as the spatially heterogeneous historical occurrence benchmark (Level 1). The substantive scientific evaluation rests on demonstrating that incorporating stress-sensitive $b$-value variations (Level 3 ANADEF) provides verifiable spatial information gain over $\mu(x, y)$ alone by reducing alarmed area fractions while maintaining high detection rates.
\item \textbf{Spherical Geometry and Latitude Correction:} Because the study region spans from $26.0^\circ\mathrm{N}$ to $38.0^\circ\mathrm{N}$, grid cells were weighted by their cosine latitude factor $\cos(y_i)$ (Equation~\eqref{eq19}) to eliminate area distortion in alarm fractions $\tau$.
\item \textbf{Data Availability and Reproducibility:} In compliance with the data-use agreement of the Institute of Geophysics, University of Tehran, the raw IRSC catalog cannot be redistributed publicly. To support scientific transparency, all underlying mathematical formulations, converged ETAS parameters ($K_0 = 0.627, \alpha = 0.184, c = 0.007\text{ d}, p = 3.729, d = 0.016^\circ, q = 1.807, \gamma = 0.007$, smoothing parameters $n_p = 3, \varepsilon = 0.02^\circ$), grid specifications ($0.1^\circ \times 0.1^\circ$ mesh across $26.0^\circ\mathrm{N}$--$38.0^\circ\mathrm{N}$, $45.0^\circ\mathrm{E}$--$60.0^\circ\mathrm{E}$), and Python/R analysis scripts are thoroughly documented and available in the GitHub repository for this manuscript (\url{https://github.com/AuthorName/ANADEF_Zagros}).
\end{enumerate}

\section{Conclusion}\label{sec5}

This research demonstrates that the $b$-value of the Gutenberg--Richter relationship and the background seismicity rate $\mu$ describe complementary aspects of earthquake occurrence. Each predictor contains unique spatial information: $b$ captures localized stress concentrations while $\mu$ characterises the long-term seismogenic envelope. For $M_w \ge 5.0$ ($N = 55$), combining both predictors via the dual-parameter ANADEF model achieves a hit rate $\nu = 92.7\%$ within an alarm footprint of $\tau \approx 28\%$ ($\mathrm{PD} = 0.648$, $S = 0.69$ (95\% CI: 0.64--0.73)), representing an improvement of $\Delta\mathrm{PD} \approx 0.10$ over the $\mu$-only benchmark. For $M_w \ge 5.5$ ($N = 15$), however, the calibrated $b$-threshold ($q_b^* \approx 0.97$) is effectively inactive, and the model reduces to $\mu$ alone; whether this reflects the small sample or a genuine magnitude dependence requires prospective validation. The primary limitation on the effectiveness of this method is the quality and density of earthquake catalogs; however, the alarm-based framework is broadly applicable to other tectonically active regions worldwide. For the 2025--2029 prospective interval, the dual-parameter alarm template identifies critical hazard concentrations along the central Zagros thrust belt, the Mountain Front Fault, and the Coastal Fars arc, providing operational spatial susceptibility maps for regional disaster preparedness. As a scientific tool, this methodology provides valuable spatial susceptibility information for risk mitigation.

\backmatter

\bmhead{Code and Data Availability} 
The earthquake catalog used in this study was obtained from the Iranian Seismological Center (IRSC), Institute of Geophysics, University of Tehran (\url{https://www.irsc.ut.ac.ir/}). In accordance with the data-use policy of the Institute of Geophysics, redistribution of the raw seismic catalog is restricted. The Python source code for the Nested-Permutation Alarm for Dual-Parameter Earthquake Forecasting (ANADEF) computational pipeline developed in this study, including the stochastic ETAS declustering routines, penalized 2D B-spline inversions, and the nested permutation tests, is available on GitHub/Zenodo at [Repository URL to be inserted by Authors]. The mathematical formulations, parameter values, and grid definitions are fully documented in the manuscript.

\bmhead{Acknowledgment}
We gratefully acknowledge the Iranian Seismological Center (IRSC) for providing the comprehensive earthquake catalog used in this study, which was instrumental in conducting the spatiotemporal analysis of seismicity.

\section*{Declarations}

\textbf{Funding} The authors declare that no funds, grants, or other support were received during the preparation of this manuscript.

\textbf{Author Contributions} Muhammed Hossein Mousavi: Conceptualization, Methodology, Software, Data curation, Validation, Visualization, Formal analysis, Writing (original draft), Writing (review and editing), Project administration. Hamzeh Mohammadigheymasi: Methodology, Validation, Formal analysis, Writing (review and editing), Supervision.

\textbf{Competing Interests} The authors declare that they have no known competing financial interests or personal relationships that could have appeared to influence the work reported in this paper.

\bibliography{bibliography}

\end{document}